\documentclass[prd,preprint,
superscriptaddress,showpacs,nofootinbib,%
tightenlines ]{revtex4}
\usepackage{epsfig}
\usepackage{amssymb}

\newcommand{\ben}{\begin{displaymath}}
\newcommand{\een}{\end{displaymath}}
\newcommand{\be}{\begin{equation}}
\newcommand{\ee}{\end{equation}}
\newcommand{\bea}{\begin{eqnarray}}
\newcommand{\eea}{\end{eqnarray}}
\newcommand{\nn}{\nonumber \\ }

\usepackage{color}

\begin{document}
\title{Wilsonian renormalization group versus subtractive\\
renormalization in effective field theories\\
for nucleon--nucleon scattering}
\author{Evgeny~Epelbaum}
 \affiliation{Institut f\"ur Theoretische Physik II, Ruhr-Universit\"at Bochum,  D-44780 Bochum,
 Germany}
\author{Jambul~Gegelia}
\affiliation{Institute for Advanced Simulation, Institut f\"ur Kernphysik
   and J\"ulich Center for Hadron Physics, Forschungszentrum J\"ulich, D-52425 J\"ulich,
Germany}
\affiliation{Tbilisi State  University,  0186 Tbilisi,
 Georgia}
\author{Ulf-G.~Mei\ss ner}
 \affiliation{Helmholtz Institut f\"ur Strahlen- und Kernphysik and Bethe
   Center for Theoretical Physics, Universit\"at Bonn, D-53115 Bonn, Germany}
 \affiliation{Institute for Advanced Simulation, Institut f\"ur Kernphysik
   and J\"ulich Center for Hadron Physics, Forschungszentrum J\"ulich, D-52425 J\"ulich,
Germany}

\begin{abstract}
We compare the subtractive renormalization and the Wilsonian
renormalization group approaches in the context of an effective field
theory for the two-nucleon system. Based on an exactly solvable model
of contact interactions, we observe that the standard Wilsonian
renormalization group approach with a single cutoff parameter does
not cover the whole space spanned by the renormalization scale
parameters of the subtractive formalism. In particular,
renormalization schemes corresponding to Weinberg's power counting
in the case of an
unnaturally large scattering length are beyond the region covered by
the Wilsonian renormalization group approach. In the framework of pionless
effective field theory, also extended by the inclusion of a long-range
interaction of separable type, we demonstrate that Weinberg's power
counting scheme is consistent in the sense that it leads to a
systematic order-by-order expansion of the scattering
amplitude.
\end{abstract}

\maketitle

\section{Introduction}
\label{introduction}

More than two decades after the publication of the ground-breaking
papers by Weinberg on the chiral effective field theory (EFT) approach
to few-nucleon systems~\cite{Weinberg:1990rz,Weinberg:1991um}, the problem of
renormalization and power counting within this formalism
still remains a hotly debated issue. The main difficulty with
Weinberg's scheme is related to the fact that
the truncated nucleon-nucleon potential within the Lippmann-Schwinger
(LS) equation is not renormalizable. Iterations of the integral
equation generate
ultraviolet (UV) divergences which cannot be absorbed by renormalizing
the parameters of the truncated potential.
An infinite number of counter terms are needed already at leading
order (LO) to cancel
UV divergences in iterations of the one-pion exchange (OPE)
potential \cite{Savage:1998vh}. In certain cases it
is possible to obtain finite, cutoff independent results from the LS
equation by taking the cutoff $\Lambda$ to infinity (or, equivalently,
much larger than all scales of the considered problem)
non-perturbatively while the perturbative iterations remain divergent,
see e.g. Refs.~\cite{Nogga:2005hy,PavonValderrama:2005gu}.
However, in EFT, all UV divergences emerging from iterations of the LS
equation should be absorbed by counter terms \cite{Epelbaum:2009sd}.

The above-mentioned renormalization problem can be avoided by treating
the pion exchange contributions to the potential perturbatively
as proposed by Kaplan, Savage and Wise \cite{Kaplan:1998tg}.
Their approach makes use of dimensional regularization supplemented
by the power divergence subtraction scheme. It leads to  the power
counting scheme, which is commonly referred to as the KSW counting.
However, it turned out that the perturbative series do not converge
within this approach (at least) in certain spin-triplet channels
\cite{Gegelia:1998ee,Cohen:1998jr,Gegelia:1999ja,Fleming:1999ee}.

Plenty of alternative, sometimes contradicting each other, formulations of the
chiral EFT in the few-nucleon sector have been suggested and/or are being
explored
\cite{Nogga:2005hy,PavonValderrama:2005gu,Epelbaum:2009sd,Kaplan:1998tg,
Ordonez:1995rz,Lepage:1997cs,Gegelia:1998iu,Gegelia:2004pz,Epelbaum:2006pt,
Mondejar:2006yu,Long:2007vp,Yang:2009pn,Birse:2010fj,Valderrama:2009ei,
Valderrama:2011mv,Long:2011xw,Long:2012ve,Long:2013cya,Beane:2008bt,
Gasparyan:2012km,Gasparyan:2013ota,Epelbaum:2012ua,Epelbaum:2012cv,
Epelbaum:2013ij,Epelbaum:2013fwa,Epelbaum:2013naa,Harada:2010ba,
Harada:2013hga,Birse:1998dk,Epelbaum:2014efa,Epelbaum:2014sza,
Gegelia:2001ev,Gegelia:1998gn,PavonValderrama:2005ku,Soto:2007pg,Soto:2009xy,
PavonValderrama:2016lqn}, see
Refs.~\cite{Bedaque:2002mn,Epelbaum:2008ga,Epelbaum:2012vx,Machleidt:2011zz})
for review articles. However, it must be said that the original Weinberg
approach using a finite momentum- or coordinate-space regulator (or
combinations thereof) is extremely successful and remains to be the
most widely employed framework. It is the aim of this paper to
demonstrate that the often made statement, that the Weinberg scheme is
inconsistent and thus its phenomenological success appears unfounded,
is misleading.

In the current work we consider exactly renormalizable
models of the nucleon-nucleon (NN) potential
which lead to well-defined scattering amplitudes. This allows us  to
compare subtractive renormalization to the Wilsonian RG approach.
We show that certain statements made in the literature are not
correct. In particular, by exploring the full space of subtractive renormalization,
we explicitly demonstrate that both the KSW and Weinberg approaches
are consistent schemes and allow for a systematic expansion of the
scattering amplitude. This can be understood by performing the
subtractions at various scales underlying the corresponding EFTs. In
fact, these two approaches simply correspond to two particular
choices of the renormalization conditions. Earlier approaches making
use of the Wilsonian renormalization group (RG) did not employ this freedom.
Thus, while these earlier findings are certainly valid, they do not
correspond to the most general situation as considered here.
We also argue that the frequently made statement that the Weinberg
approach corresponds to the expansion about the  trivial
fixed point is not correct.

Our paper is organized as follows. We begin in Section~\ref
{sec:VTcount} with general comments on the unavoidable
ambiguities when setting up the power counting schemes for NN
scattering with an unnaturally large scattering length. We show qualitatively
how the KSW and Weinberg power counting schemes emerge by employing
specific counting rules for the building blocks entering the
Lippmann-Schwinger equation for the scattering amplitude.
Next, in Sec.~\ref{sec:pionless}, we introduce an exactly solvable,
renormalizable model of contact interactions and give results for the
scattering amplitude using the most general subtractive
renormalization scheme. In Sec.~\ref{sec:wilson} we perform  the
Wilsonian RG analysis of our model following the approach
 of Ref.~\cite{Birse:1998dk} (see also Ref.~\cite{Valderrama:2016koj}
 for a recent review and related discussion) and compare the results
with  both the  Weinberg and the KSW schemes. In
section~\ref{sec:long}, we extend our analysis by considering another
exactly solvable and renormalizable model of NN scattering which
features a  long-range  interaction of a separable type.
The main results of our work are summarized in Sec.~\ref{sec:summ}
while  various lengthy expressions can be found in the appendix.

\section{On the power counting for the scattering amplitude and the potential}
\label{sec:VTcount}

The goal of any EFT is to provide an expansion of observables in powers of ratios of
small scales divided by large scales. In this section, we consider  an  EFT of NN
scattering at very low energies. The starting point is a  LS
equation build upon a two-nucleon potential. We now comment
on the relation
between the expansion of the scattering amplitude and the expansion of the potential
by taking the expansion of the amplitude as an input, fixed by the underlying theory in
terms of the low-energy scattering parameters (scattering length, effective range, $\ldots$).

To be specific, consider the $^1S_0$ partial wave of NN scattering in
non-relativistic EFT with nucleons alone as
dynamical degrees of freedom. The inverse scattering length and the
three-momenta of  the incoming and outgoing
nucleons in center-of-mass frame are the only small scales of the
considered problem, collectively denoted by $Q_S$.
At low energies, the $^1S_0$ partial wave NN scattering amplitude can
be written as a perturbative series
corresponding to the effective range expansion (ERE)
\begin{eqnarray}
T & = & T_{-1}+T_0+T_1 + \cdots, \nonumber\\
T_{-1} & = & -\frac{4\,\pi}{m_N \left(-1/a - i\, k\right)} ,\nonumber\\
T_0    & = & \frac{2\,\pi \, r_e k^2}{m_N \left(-1/a - i\, k\right)^2} ,\nonumber\\
T_1    & = & -\frac{\pi  \,r_e^2 k^4}{m_N
   (-1/a-i\,k)^3}, \nonumber\\
       & \cdots & ,
\label{Texp}
\end{eqnarray}
where the subscripts indicate the orders in the small parameter $Q_S$,
while $a$ and $r_e$ refer to the scattering length and
the effective range, respectively. We take this sequence of approximations to the amplitude
and demand that it is reproduced order-by-order by the  low-energy EFT.
Note that while the LO potential has to be iterated to reproduce the
LO amplitude, higher order corrections can be included
perturbatively. For the case of contact interactions the
amplitude can be calculated analytically and the renormalization can
be carried out explicitly. Therefore, it can be shown explicitly that
the non-perturbative and perturbative treatments of higher order
corrections to the potential differ by higher order contributions
which are beyond the accuracy of the calculation (see
e.g.~Refs.~\cite{Gegelia:1998iu,Gegelia:1998gn}). These formally
higher order corrections are indeed small provided that the
renormalization conditions are appropriately chosen as will be
discussed in detail below.

We now make the connection to an underlying NN potential. To assign
orders of the small parameter to the various terms in the effective
potential
corresponding to Eq.~(\ref{Texp}) we write it as a perturbative series
\begin{equation}
V =   V_{\rm LO} + V_{\rm NLO}+ V_{\rm NNLO}  + \cdots ,
\label{Vexp}
\end{equation}
where the orders corresponding to the different terms need to be
obtained by analyzing the integral equation for the scattering
amplitude.
Each next term in the sequence $V_{\rm LO}, \  V_{\rm NLO}, \ V_{\rm
  NNLO}, \ \cdots$  is suppressed by some power of the small parameter
compared to the previous term.

To obtain the leading order amplitude $T_{-1}$ we need to solve the LS equation:
\begin{equation}
T_{-1} =  V_{\rm LO}+V_{\rm LO}\, G\, T_{-1},
\label{teq}
\end{equation}
where $G$ is the resolvent--operator of the two--nucleon propagator.
The solution to this equation has the form:
\begin{equation}
T_{-1}=(1-V_{\rm LO}\, G )^{-1} V_{\rm LO}.
\label{ft2}
\end{equation}
Eq.~(\ref{ft2}) can be satisfied by assigning the orders as follows:
\begin{eqnarray}
V_{LO} &\sim& \epsilon ^x,\nonumber\\
1-V_{LO}\,G&\sim& \epsilon^{1+x},\nonumber\\
G &\sim& \epsilon^{-x} ({\rm or} \ \epsilon^1, \ {\rm if} \ \ x\leq -1 ),
\label{ft3}
\end{eqnarray}
where $\epsilon \sim Q_S/\Lambda_H$, with $\Lambda_H$
denoting the hard (breakdown) scale. In the framework of EFT,
the order of the potential $V_{\rm LO}$ depends on the choice of the
renormalization condition.
The choice $x=0$ corresponds to Weinberg's power counting
\cite{Weinberg:1990rz} for the NN potential in EFT.
The KSW counting \cite{Kaplan:1998tg} for the LO potential is the
realization of the choice $x=-1$,
the same counting for the LO potential is advocated by the
renormalization group approach of Ref.~\cite{Birse:1998dk}.

Next, let us  investigate the power counting for the higher
order contributions in the potential for both choices $x=0$ and $x=-1$.
For that we write the amplitude $T_0$ as
\begin{equation}
T_0 = V_{\rm NLO}+V_{\rm NLO} \,G \, T_{-1}+T_{-1} \, G \, V_{\rm NLO}
+T_{-1}\,G\,V_{\rm NLO} \, G \,  T_{-1} \, .
\label{ttildeeq}
\end{equation}
It follows from Eq.~(\ref{ttildeeq}) that $V_{\rm NLO}\sim \epsilon^0$
for $x=-1$, and $V_{NLO}\sim \epsilon^2$ for $x=0$.
Higher order terms in the potential can be analyzed analogously.

Thus, we observe that the proper scaling of the physical amplitude in terms of the small parameter can be
realized by a potential, whose various contributions behave as
\begin{equation}
V_{\rm LO}  \sim  \epsilon^0 ,\quad \quad
V_{\rm NLO}  \sim  \epsilon^2, \quad \quad
V_{\rm NNLO}  \sim  \epsilon^4, \quad \quad
 \cdots ,
\label{xeq0}
\end{equation}
or
\begin{equation}
V_{\rm LO}  \sim  \epsilon^{-1} ,\quad \quad
V_{\rm NLO}  \sim  \epsilon^0, \quad \quad
V_{\rm NNLO}  \sim  \epsilon^2, \quad \quad
 \cdots
\label{xm1}
\end{equation}
depending on the employed choice of the renormalization conditions.

To compare the counting rules for the various terms
with the actual power counting of the potential obtained in pionless EFT
consider the EFT potential for the $^1S_0$ partial wave
\begin{equation}
V =c+c_2 \left(p^2+p'^2\right) +c_4(p^4+p'^4)+c_{22}\, p^2 p'^2 , \label{potentialCI}
\end{equation}
where, for simplicity, we set $c_{22}=0$.
The solution to the LS equation using dimensional regularization has the form
\begin{equation}
T(k)=\frac{c+2  c_2 k^2
\left(c_2+c_4  k^2\right)
}{1-I\left(k^2\right) \left[c+2  c_2 k^2
   \left(c_2+c_4 k^2\right)
\right]}.
\label{TMat}
\end{equation}
Subtracting the loop integral $I(k^2)=m \sqrt{-k^2-i\epsilon}/(4 \pi)$ at $k^2=-\mu^2$ and matching to the ERE one obtains:
\begin{eqnarray}
\frac{1}{T(k)}&=& \frac{m}{4 \pi } \left[\frac{4 (1-a \mu )^3}{a^2 k^2 r_e
   \left(a k^2 r_e-2 a \mu +2\right)   -4 a^2 k^4 v_2 (a \mu -1)
   +4 a
   (a \mu -1)^2}+\mu +i \,k\right],\nonumber\\
c & = & \frac{4 \pi }{m(1/a- \mu )},\nonumber\\
c_2 & = & \frac{\pi  r_e}{m(1/a-  \mu )^2}.
\nonumber\\
c_4 & = & \frac{\pi  r_e^2}{2 m (1/a -  \mu )^3}+\frac{2 \pi  v_2}{m (1/a -  \mu )^2}.
\label{fits}
\end{eqnarray}
By choosing $\mu$ of the order of the hard scale $\Lambda_H$ and taking into account that
$a\sim \epsilon^{-1}$, $ r_e\sim v_2\sim \epsilon^0$, we see from Eq.~(\ref{fits}) that the
coupling constants are of a natural size, $c_i \sim \epsilon^0$,
leading to the scaling of various terms in the potential according to
Eq.~(\ref{xeq0}). That is, for this choice of renormalization
conditions
Eq.~(\ref{xeq0}) corresponds to Weinberg's power counting. On the other hand, if we take
$\mu$ of the order of the small scale $Q_S$, the couplings $c$, $c_2$
and $c_4$ comply with
the KSW
counting.

It is important to emphasize that a certain amount of fine tuning
in the scattering amplitude  beyond naive dimensional analysis
is unavoidable in the case of
an unnaturally large scattering length both for the Weinberg and KSW
power countings. In the Weinberg case, the fine tuning manifests itself in
the second condition in Eq.~(\ref{ft3}). For the KSW counting, one
observes that the constant $c_4$ actually  violates the scaling suggested by
Eq.~(\ref{xm1}). That is, in the KSW counting, the scaling of the
amplitude comes out as a result of cancelation of large
contributions between the $c_2^2$ and $c_4$  contributions, which goes
beyond naive dimensional analysis.

To demonstrate in more details the  above observations and to compare to the Wilsonian
RG approach we now consider solvable toy models of the NN
interaction. These models are exactly
renormalizable and hence demonstrate an essential feature of consistent EFTs
that a perturbative expansion of renormalized non-perturbative expressions, if expanded,
reproduce the standard renormalized perturbative series.

\section{An exactly solvable model of contact interactions}
\label{sec:pionless}

Consider an exactly solvable model of the fully off-shell LS equation
\begin{equation}
T(p',p,k) = V(p',p,k) + 2 m \int \frac{d^3l}{(2
\pi)^3}\,V(p',l,k)\,\frac{1}{k^2-l^2+i\,\eta}\, T(l,p,k)
\label{LSEquation}
\end{equation}
with the potential
\begin{equation}
V(p',p,k) = \left(1, p'^2\right)\,\lambda(k)\,\left(
\begin{array}{l}
 1 \\
 p^2
\end{array}
\right), \label{potential}
\end{equation}
where $\lambda$ is a $2\times 2$ matrix given by
\begin{equation}
\lambda(k) = \left(
\begin{array}{ll}
C+k^2 C_{\rm k}(k^2) & C_2 + k^2\,C_{\rm 2E} + k^4 C_{\rm 2 k}(k^2) \\
C_2 + k^2\,C_{\rm 2E}  + k^4 C_{\rm 2 k}(k^2) & C_4 +k^2 \,C_{\rm 4E} + k^4
\,C_{\rm 4EE} +k^6\,C_{\rm 4 k}(k^2)
\end{array}
\right)^{-1}. \label{lambdaminus1}
\end{equation}
Here, $C_{\rm k}(k^2)$, $C_{\rm 2 k}(k^2)$ and $C_{\rm 4 k}(k^2)$ are analytic
functions of $k^2$ at $k^2=0$, i.e. they can be expanded in a Taylor series
in  $k^2$.

By writing
\begin{equation}
T(p',p,k) = \left(1, p'^2\right)\,\tau(k)\,\left(
\begin{array}{l}
 1 \\
 p^2
\end{array}
\right) \label{amplpar}
\end{equation}
Eq.~(\ref{LSEquation}) is reduced to a matrix equation
\cite{Phillips:1997xu}
\begin{equation}
\tau(k) = \lambda(k)+\lambda(k){\cal G}(k)\tau(k)\,,
\label{matrixeq}
\end{equation}
with
\begin{equation}
{\cal G}(k) =  \left(
\begin{array}{ll}
I^\Lambda(k) &
I^\Lambda(k) k^2+I_3^\Lambda \\
I^\Lambda(k) k^2+I_3^\Lambda & I^\Lambda(k) k^4+I_3^\Lambda
k^2+I_5^\Lambda
\end{array}
\right). \label{gmat}
\end{equation}
The cutoff-regularized loop integrals of Eq.~(\ref{gmat}) are defined as
\begin{eqnarray}
I_n^\Lambda & = & -{m \over (2\pi )^3}\int {d^3l}\, l^{n-3}\,\theta
\left(\Lambda-l \right) = -\frac{m\,\Lambda^n}{2n\pi^2} \,,\; \mbox{
  with } \; n=1,3,5\,, \nn
I^\Lambda(k) & = & {m\over (2\pi )^3}\int {d^3l \,\theta
\left(\Lambda-l
  \right)\over
k^2-l^2+i\eta} = I_1^\Lambda -\frac{i\,m\, k}{4\pi} - \frac{m
k}{4\pi^2} \,\ln \frac{\Lambda-k}{\Lambda+k}\,,
\label{3}
\end{eqnarray}
where the last equation is valid for $k < \Lambda$.
By writing the matrix equation (\ref{matrixeq}) as
\begin{equation}
\tau(k)^{-1} = \lambda(k)^{-1}-{\cal G}(k)\,, \label{matrixeqinv}
\end{equation}
one can easily see from Eqs.~(\ref{lambdaminus1}), (\ref{gmat}) and (\ref{3}),
that the whole $\Lambda$-dependence present in ${\cal G}(k)$ can be
eliminated by choosing
\begin{eqnarray}
C & = & \alpha -\frac{ m \Lambda }{2 \pi ^2}\equiv -\frac{ m
\Lambda }{2 \pi ^2}+\frac{ m \mu }{2 \pi ^2}+C_{\rm R}(\mu
   ),\nonumber\\
C_2 & = & \beta -\frac{ m \Lambda ^3}{6 \pi ^2} \equiv -\frac{ m
\Lambda
   ^3}{6 \pi ^2}+\frac{ m
\mu_1^3}{6 \pi
   ^2}+C_{\rm 2R}(\mu_1),\nonumber\\
C_4 & = & \gamma -\frac{ m \Lambda ^5}{10 \pi ^2} \equiv -\frac{
m \Lambda ^5}{10 \pi ^2} +\frac{ m \mu_3^5}{10 \pi
   ^2}+C_{\rm 4R}(\mu_3),\nonumber\\
C_{\rm 2E} & = & \delta -\frac{ m \Lambda }{2\,\pi ^2} \equiv
-\frac{ m \Lambda}{2 \pi
   ^2}+\frac{ m \mu_2}{2 \pi
   ^2}+C_{\rm 2ER}(\mu_2),\nonumber\\
C_{\rm 4E} & = & \lambda -\frac{ m \Lambda ^3}{6 \pi ^2}\equiv
-\frac{ m \Lambda
   ^3}{6 \pi ^2}+
\frac{ m \mu_4^3}{6 \pi
   ^2}+C_{\rm 4ER}(\mu_4),\nonumber\\
C_{\rm 4EE} & = & \sigma -\frac{\, m \Lambda }{2 \pi ^2}\equiv
-\frac{ m \Lambda}{2 \pi
   ^2}+\frac{ m \mu_5}{2 \pi
   ^2}+C_{\rm 4EER}(\mu_5),\nonumber\\
C_{\rm k}(k^2) & = & C_{\rm P}(k^2)-\frac{ m \ln \frac{\Lambda
-k}{k+\Lambda }}{4 k \pi
   ^2}\,,\nonumber\\
C_{\rm 2 k}(k^2) & = & C_{\rm 2P}(k^2)-\frac{ m \ln \frac{\Lambda
-k}{k+\Lambda }}{4 k \pi
   ^2}\,,\nonumber\\
C_{\rm 4 k}(k^2) & = & C_{\rm 4P}(k^2)-\frac{ m \ln \frac{\Lambda
-k}{k+\Lambda }}{4 k \pi
   ^2}\,,
    \label{bareC}
\end{eqnarray}
where $\alpha$, $\beta$, $\gamma$, $\delta$, $\lambda$ and $\sigma$
are some finite parameters. We have introduced the scale-dependent renormalized
coupling constants $C_{\rm R}(\mu)$, $C_{\rm 2R}(\mu_1)$, etc., where  $\mu$, $\mu_1,\ldots,\mu_5$
are renormalization scale parameters and $C_{\rm P}(k^2)$, $C_{\rm 2P}(k^2)$,
$C_{\rm 4P}(k^2)$ are finite functions of $k^2$, analytic  at $k^2=0$.
We set these functions equal to zero for the sake of simplicity.
Substituting Eq.~(\ref{bareC}) into the solution of Eq.~(\ref{LSEquation})
we obtain the expression for the inverse of the on-shell scattering amplitude
\begin{eqnarray}
\frac{1}{T(k)} & = &  \frac{N_R}{D_R}+  \frac{i  k\,m}{4 \pi } \nonumber\\
&\equiv& -\frac{k^4 (\delta ^2-\alpha  \sigma) +k^2 (2 \beta  \delta - \alpha \lambda) +\beta ^2 - \alpha\gamma }{k^4 (\alpha -2
   \delta +\sigma )+k^2 (\lambda -2 \beta )+\gamma }
 +  \frac{i  k\,m}{4 \pi }\,,
 \label{rampl}
 \end{eqnarray}
 where
 \begin{eqnarray}
 N_R & = & 6 \pi ^2 \biggl\{
 C_{\rm R}(\mu ) \biggl[
 30 \pi ^2 \left(\text{\textit{$k$}}^4
   C_{4 {\rm EER}}\left(\text{\textit{$\mu
   $}}_5\right)+\text{\textit{$k$}}^2 C_{4
   {\rm ER}}\left(\text{\textit{$\mu $}}_4\right)+C_{\rm 4
   R}\left(\text{\textit{$\mu $}}_3\right)\right) \nonumber\\
   &+& m \left(5
   \text{\textit{$k$}}^2 \left(3 \text{\textit{$k$}}^2
   \text{\textit{$\mu $}}_5+\text{\textit{$\mu $}}_4^3\right)+3
   \text{\textit{$\mu $}}_3^5\right)
   \biggr]
%   \nonumber\\
  -  5 \biggl[
  -3 m \mu
   \left(\text{\textit{$k$}}^4 C_{\rm 4 {EER}}\left(\text{\textit{$\mu
   $}}_5\right)+\text{\textit{$k$}}^2 C_{\rm 4
   {ER}}\left(\text{\textit{$\mu $}}_4\right)+C_{\rm 4
   R}\left(\text{\textit{$\mu $}}_3\right)\right) \nonumber\\
   &+& 2
   \text{\textit{$k$}}^2 C_{\rm 2 {ER}}\left(\text{\textit{$\mu
   $}}_2\right) \left(6 \pi ^2 C_{\rm 2 R}\left(\text{\textit{$\mu
   $}}_1\right)+m \left(3 \text{\textit{$k$}}^2 \text{\textit{$\mu
   $}}_2+\text{\textit{$\mu $}}_1^3\right)\right)+6 \pi ^2
   \text{\textit{$k$}}^4 C_{\rm 2 {ER}}\left(\text{\textit{$\mu
   $}}_2\right){}^2 \nonumber\\
   &+&  2 m \left(3 \text{\textit{$k$}}^2
   \text{\textit{$\mu $}}_2+\text{\textit{$\mu $}}_1^3\right) C_{\rm 2
   R}\left(\text{\textit{$\mu $}}_1\right)+6 \pi ^2 C_{\rm 2
   R}\left(\text{\textit{$\mu $}}_1\right){}^2
   \biggr]
   \biggr\}
   \nonumber\\
   &+& m^2
   \left[9 \mu  \text{\textit{$\mu $}}_3^5-5 \left(-3 \mu
   \text{\textit{$k$}}^2 \text{\textit{$\mu $}}_4^3+9
   \text{\textit{$k$}}^4 \left(\text{\textit{$\mu $}}_2^2-\mu
   \text{\textit{$\mu $}}_5\right)+6 \text{\textit{$k$}}^2
   \text{\textit{$\mu $}}_2 \text{\textit{$\mu
   $}}_1^3+\text{\textit{$\mu $}}_1^6\right)\right],
   \nonumber\\
 D_R & =& 6 \pi ^2 \biggl\{ 30 \pi ^2
 \biggl[
 {\textit{$k$}}^2
   \left({\textit{$k$}}^2 \left(C_{\rm 4
   {EER}}\left({\textit{$\mu $}}_5\right)-2 C_{\rm 2
   {ER}}\left({\textit{$\mu $}}_2\right)+C_{\rm R}(\mu
   )\right)+C_{\rm 4 {ER}}\left({\textit{$\mu $}}_4\right)-2 C_{\rm 2
   R}\left({\textit{$\mu $}}_1\right)\right)\nonumber\\
   &+& C_{\rm 4
   R}\left({\textit{$\mu $}}_3\right)
   \biggr]
   +m \left(5
   \text{\textit{$k$}}^2 \left(3 \text{\textit{$k$}}^2 \left(-2
   \text{\textit{$\mu $}}_2+\text{\textit{$\mu $}}_5+\mu \right)-2
   \text{\textit{$\mu $}}_1^3+\text{\textit{$\mu $}}_4^3\right)+3
   \text{\textit{$\mu $}}_3^5\right) \biggr\} .
 \label{ramplBBs}
\end{eqnarray}
As expected, the full scattering amplitude does not depend on the renormalization scale parameters
$\mu$, $\mu_1,\ldots,\mu_5$ as the explicit and  implicit (through the renormalized
couplings) scale dependence cancels exactly so that each of these parameters can be chosen arbitrarily.
On the other hand, the freedom of this choice can be advantageously exploited to obtain a better convergent
perturbative series when the expansion is performed in terms of the renormalized coupling(s).

It is convenient to parameterize the model in terms of
the standard EFT
expansion of the effective potential. To this end, we expand $V(p',p,k)$ in
Taylor series in $p$, $p'$ and $k$ and match
the terms up to fourth order to
\begin{equation}
V\left(p',p,k\right)  =c + c_2 (p'^2+p^2) + c_{\rm E} k^2+ c_{\rm pp} p'^2 p^2 + c_{\rm Ep} k^2 (p'^2+p^2) + c_{\rm EE} k^4 + \cdots\,.
\end{equation}
Solving the resulting system of equations we obtain the following relations
between the bare parameters
\begin{eqnarray}
C & = &  \frac{c_{\rm pp}}{c c_{\rm pp}-c_2^2}\,,\nonumber\\
C_2 & = & \frac{c_2}{c_2^2-c c_{\rm pp}}\,,\nonumber\\
C_4 & = & \frac{c}{c c_{\rm pp}-c_2^2}\,,\nonumber\\
C_{\rm 2E} & = & \frac{c_{\rm E} c_{\rm pp}-c_2 c_{\rm Ep}}{c_2^3-c
   c_2 c_{\rm pp}}\,,\nonumber\\
C_{\rm 4E} & = & -\frac{c_{\rm E} c_2^2-2 c c_{\rm Ep} c_2+c
   c_{\rm E} c_{\rm pp}}{c_2^4-c c_2^2 c_{\rm pp}}\,,\nonumber\\
C_{\rm 4EE} & = &  -\frac{c_{\rm EE} c_2^2-2 c_{\rm E} c_{\rm Ep} c_2+c
   c_{\rm Ep}^2+c_{\rm E}^2 c_{\rm pp}-c c_{\rm EE}
   c_{\rm pp}}{c_2^4-c c_2^2 c_{\rm pp}}\,\, .
 \label{reparametrizing}
\end{eqnarray}
Thus, the Taylor expansion of the potential in powers of momenta and energy in
this new parametrization has the form analogous to the one of the EFT potential
with contact interactions.

Substituting the bare couplings of the new parametrization expressed in terms of
renormalized ones as specified in the appendix
into the solution to the LS equation we obtain the following renormalized expression
for the amplitude:
\begin{eqnarray}
T(k) & = & \frac{D}{N+\frac{i  k\,m}{4 \pi } D} \,,\nonumber\\
N & = & 90 \pi ^2 k^2 \biggl[k^2 \left(c_{\text{EpR}}^2-c_{\text{EER}}
   c_{\text{ppR}}\right) \left(m \mu  c_{\rm R}+2 \pi ^2\right)+k^2 m
   \mu  c_{\text{ER}}^2 c_{\text{ppR}}\nonumber\\
&+& c_{\text{ER}}
   c_{\text{ppR}} \left(m \mu  c_{\rm R}+2 \pi ^2\right)\biggr]
   +c_{\rm 2 R}^2 \biggl\{ 6 \pi ^2 m \biggl[5 k^2 \biggl(3 k^2 \left(\mu
   c_{\text{EER}}-2 \mu_2 c_{\text{EpR}}+\mu_5
   c_{\text{ppR}}\right) \nonumber\\
&-& 2 \mu_1^3 c_{\text{EpR}}+3 \mu
    c_{\text{ER}}+\mu_4^3 c_{\text{ppR}}\biggr)+3
   \mu_3^5 c_{\text{ppR}}+15 \mu  c_{\rm R}\biggr] \nonumber\\
&+ & m^2
   c_{\text{ppR}} c_{\rm R} \left(9 \mu  \mu_3^5-5 \left(9
   k^4 \left(\mu_2^2-\mu  \mu_5\right)-3 k^2
   \mu  \mu_4^3+6 k^2 \mu_1^3 \mu_2+\mu_1^6\right)\right)+180 \pi ^4\biggr\}\nonumber\\
&- & 60 \pi
   ^2 k^2 c_{\rm 2 R} \biggl[3 c_{\text{EpR}} \left(k^2 m \mu
   c_{\text{ER}}+m \mu  c_{\rm R}+2 \pi ^2\right)-m c_{\text{ER}}
   c_{\text{ppR}} \left(3 k^2 \mu_2+\mu_1^3\right)\biggr]\nonumber\\
&+& 60 \pi ^2 m c_{\rm 2 R}^3 \left(3 k^2
   \mu_2+\mu_1^3\right)+m^2 c_{\rm 2 R}^4 \biggl[5
   \biggl(9 k^4 \left(\mu_2^2-\mu  \mu_5\right)-3 k^2 \mu  \mu_4^3 \nonumber\\
&+& 6 k^2 \mu_1^3
   \mu_2+\mu_1^6\biggr)-9 \mu  \mu_3^5\biggr]   \,,\nonumber\\
D & = & 6 \pi ^2 \biggl\{ c_{\rm 2 R}^2 \biggl[30 \pi ^2 k^2 \left(k^2
   \left(c_{\text{EER}}-2
   c_{\text{EpR}}+c_{\text{ppR}}\right)+c_{\text{ER}}\right)\nonumber\\
&+& c_{\rm R}
   \left(m c_{\text{ppR}} \left(5 k^2 \left(3 k^2 (\mu -2
   \mu_2+\mu_5)-2 \mu_1^3+\mu_4^3\right)+3 \mu_3^5\right)+30 \pi
   ^2\right)\biggr]\nonumber\\
&+& 30 \pi ^2 k^2 \biggl[k^2 \left(c_{\rm R}
   \left(c_{\text{EpR}}^2-c_{\text{EER}}
   c_{\text{ppR}}\right)+c_{\text{ER}}^2
   c_{\text{ppR}}\right)+c_{\text{ER}} c_{\text{ppR}}
   c_{\rm R}\biggr] \nonumber\\
&-& 60 \pi ^2 k^2 c_{\rm 2 R} \biggl[k^2 c_{\text{ER}}
   \left(c_{\text{EpR}}-c_{\text{ppR}}\right)+c_{\text{EpR}}
   c_{\rm R}\biggr] \nonumber\\
&-& m c_{\rm 2 R}^4 \biggl[ 5 k^2 \biggl(3 k^2 (\mu -2
   \mu_2+\mu_5) -2 \mu_1^3+\mu_4^3\biggr) +3 \mu_3^5\biggr] +60 \pi ^2 k^2 c_{\rm 2
   R}^3\biggr\} \,. \label{rentmp2}
\end{eqnarray}

For our purposes it is sufficient
%convenient
to consider a particular case by taking
\begin{equation}
\label{constraints}
\lambda=2 \beta\,, \quad \quad  \sigma = 2\delta-\alpha\,,
\end{equation}
for which the inverse amplitude reduces to
\begin{eqnarray}
\frac{1}{T(k)}  & = &  \frac{- (\alpha-\delta)^2 k^4 +2 \beta (\alpha- \delta)  k^2+\alpha\gamma -\beta ^2}{\gamma }
 +  \frac{i  k\,m}{4 \pi } \nonumber\\
 &\equiv& \frac{- ((\alpha-\delta)/\beta)^2 k^4 +2 ((\alpha- \delta)/\beta)  k^2+\alpha (\gamma/\beta^2) - 1}{(\gamma/\beta^2) }
 +  \frac{i  k\,m}{4 \pi } \,.
\label{Tch}
\end{eqnarray}
As can be seen from Eq.~(\ref{Tch}), the inverse amplitude depends only on three independent parameters
which can be conveniently expressed in terms of the scattering length $a$,
effective range $r_e$ and the first shape parameter $v_2$ via
\begin{eqnarray}
\alpha &=& \frac{m \left(a\, r_e^2+16 v_2\right)}{64 \pi  a v_2},\nonumber\\
\frac{\gamma}{\beta^2} &=& \frac{64 \pi  v_2}{m r_e^2} \,,\nonumber\\
\frac{\alpha-\delta}{\beta} &=& - \frac{4 v_2}{r_e} \,.
\label{parameters}
\end{eqnarray}
The scattering amplitude then takes the familiar form
\begin{equation}
T(k) = -\frac{4\,\pi}{m}\left[ -\frac{1}{a}+\frac{r_e k^2 }{2}+v_2 k^4 -i\,k\right]^{-1}.
\label{invamp0}
\end{equation}
Renormalized couplings corresponding to
this particular choice
are given in Eq.~(\ref{rencouplings})
of the appendix.
As mentioned above, the scattering amplitude does not depend on the
redundant constant $\beta$, which can be viewed as an off-shell
parameter in our model.
A generalization to include higher-order shape
parameters is straightforward. Notice further that
while it is perfectly fine to constrain the finite pieces of the coupling
constants  as explained above in order to simplify the analysis,
all bare parameters of the underlying model have to be taken
into account to maintain its explicit renormalizability.

To summarize, we have introduced a solvable, renormalizable model of
contact interactions specified by Eqs.~(\ref{potential}) and
(\ref{lambdaminus1}). The bare parameters of the model are
expressible in terms of the renormalized ones as given in
Eq.~(\ref{bareC}) in such a way that the iterative solution to the LS equation
(\ref{LSEquation}) remains finite in the limit of $\Lambda \to \infty$.
Notice that renormalization of the amplitude unavoidably
introduces the dependence of the renormalized couplings on the
subtraction points $\mu$, $\mu_{1}, \ldots \mu_{5}$, whose choice reflects the freedom
in the choice of renormalization conditions. Utilizing the standard
convention of pionless EFT, the parameters of our
model can be expressed in terms of the coupling constants $c$, $c_2$,
$c_{\rm E}$, $c_{\rm pp}$, $c_{\rm Ep}$ and $c_{\rm EE}$ accompanying zero-range
contact interactions. The relations between the bare and renormalized
constants in this notation are given in the appendix. Finally, to keep our analysis
simple, we restrict ourselves to a particular choice of the parameter
space of our model by setting the finite, energy-dependent functions
$C_{\rm P} (k^2)$, $C_{\rm 2P} (k^2)$ and $C_{\rm 4P} (k^2)$ equal to zero and
by constraining the parameters as specified in Eq.~(\ref{constraints}).
For this choice, the real part of the inverse scattering amplitude is
given by the first three terms of the effective range expansion.
The resulting model provides a simple framework to explore different
choices of the renormalization conditions as will be discussed in the
next section.

\section{Wilsonian Renormalization Group Analysis}
\label{sec:wilson}

We now perform a Wilsonian RG analysis of the model introduced in the
previous section along the lines of Ref.~\cite{Birse:2010fj}. For that, we introduce a scale $\Lambda$, which acts as
a cut-off on the virtual momenta, and we demand that physics does not
depend on it.  It is straightforward to check that the potential
$V(p',p,k)\equiv V(p',p,k,\Lambda)$
specified by  Eqs.~(\ref{potential}), (\ref{lambdaminus1}) and (\ref{bareC})
satisfies the RG equation \cite{Birse:1998dk}:
\begin{equation}
\frac{\partial V(p',p,k,\Lambda)}{\partial \Lambda}= \frac{m}{2
\pi^2}\,V(p',\Lambda,k,\Lambda)\,\frac{\Lambda^2}{\Lambda^2-k^2}\,V(\Lambda,p,k,\Lambda)
\, . \label{RGE1}
\end{equation}
Further, the corresponding re-scaled potential
\begin{equation}
\hat V(\hat p',\hat p,\hat k,\Lambda) :=\frac{m \Lambda}{2 \pi ^2}\,
V(\hat p' \Lambda ,\hat p \Lambda
   ,\hat k \Lambda ,\Lambda )
\label{rescaledpotential}
\end{equation}
satisfies the equation
\begin{eqnarray}
\Lambda\, \frac{\partial \hat V(\hat p',\hat p,\hat
k,\Lambda)}{\partial \Lambda} & = & \hat p'\,\frac{\partial \hat
V(\hat p',\hat p,\hat k,\Lambda)}{\partial \Lambda}+\hat
p\,\frac{\partial \hat V(\hat p',\hat p,\hat k,\Lambda)}{\partial
\Lambda}+\hat k\,\frac{\partial \hat V(\hat p',\hat p,\hat
k,\Lambda)}{\partial \Lambda}
\nonumber\\
& + & \hat V(\hat p',\hat p,\hat k,\Lambda) +\hat V(\hat p',1,\hat
k,\Lambda)\,\frac{1}{1-\hat k^2}\,\hat V(1,\hat p,\hat
k,\Lambda)\,.\label{RGE2}
\end{eqnarray}
For the choice of parameters of Eq.~(\ref{parameters}), the re-scaled potential  has the form
\begin{eqnarray}
\hat V(\hat p',\hat p,\hat k,\Lambda)
& = & \frac{\hat P_1}{\hat P_2}\,,\nonumber\\
\hat P_1 & = & 3 \Lambda  \biggl\{15 \pi  a m^2 \Lambda ^4 r_e^4
   \left(\hat p'^2-\hat k^2\right)
   \left(\hat k^2-\hat p^2\right)
\nonumber\\
&+&
 3840 \pi ^2 a m
   \hat k^2 v_2^2 \beta  \Lambda ^4 r_e
   \left(\hat p'^2+\hat p^2-2 \hat k^2\right)\nonumber\\
&+& 16 m v_2
   \Lambda ^2 r_e^2 \biggl[15 a m \hat k \Lambda ^3
   \left(\hat k^2-\hat p'^2\right)
   \left(\hat k^2-\hat p^2\right) \ln
   \frac{1-\hat k}{1+\hat k} \nonumber\\
&+& 2 a \biggl(5
   \hat p'^2 \left(m \Lambda ^3 \left(3 \hat p^2-3
   \hat k^2-1\right)+6 \pi ^2 \beta \right) \nonumber\\
&+& 5 \hat p^2
   \left(6 \pi ^2 \beta -m \left(3 \hat k^2+1\right) \Lambda
   ^3\right)+15 m \hat k^4 \Lambda ^3+5 m \hat k^2 \Lambda
   ^3+3 m \Lambda ^3\nonumber\\
&-& 60 \pi ^2 \hat k^2 \beta \biggr)+15 \pi
   m \Lambda ^2 \left(\hat p'^2-\hat k^2\right)
   \left(\hat k^2-\hat p^2\right)\biggr]-61440 \pi ^3 a
   v_2^2 \beta ^2\biggr\}\,,\nonumber\\
\hat P_2 & = & 3 \pi  a m^2 \left(3-5
   \hat k^2\right) \Lambda ^5 r_e^4+7680 \pi ^2 a \hat k^2
   v_2^2 \beta  \Lambda ^2 r_e \left(6 \pi ^2 \beta -m \Lambda
   ^3\right) \nonumber\\
&+& 16 m v_2 \Lambda ^3 r_e^2 \biggl[-2 a \left(m
   \left(4-15 \hat k^2\right) \Lambda ^3+60 \pi ^2 \beta
   \right)   \nonumber\\
&+& 3 a m \hat k \left(5 \hat k^2-3\right) \Lambda
   ^3 \ln\frac{1-\hat k}{1+\hat k}+3 \pi
   m \left(3-5 \hat k^2\right) \Lambda ^2\biggr] \nonumber\\
&+& 92160 \pi ^3
   v_2^2 \beta ^2 \left(\pi  a \hat k^4 v_2 \Lambda ^4+a
   \hat k \Lambda  \ln
   \frac{1-\hat k}{1+\hat k}+2 a \Lambda
   -\pi \right).
\label{potrescRG}
\end{eqnarray}

Following the Wilsonian RG approach of Ref.~\cite{Birse:1998dk} for
an unnaturally large scattering length,
we expand the potential $\hat V(\hat p',\hat p,\hat k,\Lambda)$ of
Eq.~(\ref{potrescRG}) in powers of $\epsilon$ by counting $a \Lambda\sim
\epsilon^0$, $\Lambda \sim \epsilon$ and obtain:
\begin{equation}
\hat V(\hat p',\hat p,\hat k,\Lambda)=\hat V_0(\hat p',\hat p,\hat k,\Lambda)+\hat V_1(\hat p',\hat p,\hat k,\Lambda) + \cdots ,
\label{Vexpanded}
\end{equation}
where
\begin{eqnarray}
\hat V_0(\hat p',\hat p,\hat k,\Lambda) &=& \frac{2}{-\hat k \, \ln
   \frac{1-\hat k}{1+\hat k}-2
   +\pi/(a\Lambda) }\sim \epsilon^0,\nonumber\\
\hat V_1(\hat p',\hat p,\hat k,\Lambda) &=& \frac{\pi  \hat k^2 \Lambda r_e}{\left(-\hat k
   \,\ln \frac{1-\hat k}{1+\hat k}-2
    +\pi/(a\Lambda) \right)^2}\sim \epsilon^1 ,\nonumber\\
&& \cdots .
\label{vtexpanded}
\end{eqnarray}
The leading order term $\hat V_0$ corresponds to a non-trivial fixed point with
an unstable perturbation
summed up to all orders \cite{Birse:1998dk}. The next term
$\hat V_1$ is of order $\epsilon^1$.  The corresponding un-scaled potentials
have the orders
\begin{eqnarray}
V_0(p', p, k,\Lambda) &=& \frac{4 \pi^2a}{m \left(\pi-2 a \Lambda-a  k \ln
   \frac{1-k/\Lambda}{1+k/\Lambda}\right) }\sim \frac{1}{\epsilon},\nonumber\\
V_1(p', p, k,\Lambda) &=& \frac{2 \pi^3  a^2 k^2 r_e}{m \left(\pi-2 a \Lambda-a  k \ln
   \frac{1-k/\Lambda}{1+k/\Lambda}\right)^2}\sim \epsilon^0 ,\nonumber\\
&& \cdots .
\label{vtexpandedunscaled}
\end{eqnarray}

This scaling behavior for the various contributions to the
potential based on the Wilsonian RG analysis can now be compared with
the one corresponding to different choices of renormalization
conditions in the subtractive approach.
Notice, however, that the potential cannot be fully determined by reproducing
the on-shell scattering amplitude. Thus, we also need to make a choice
for  the off-shell parameter $\beta$. Taking $\beta \sim \epsilon^0$
and choosing the subtraction points as
\begin{equation}
\mu \sim \mu_{1, \ldots , 5} \sim \epsilon \,,
\end{equation}
the renormalized potential takes the form
\begin{eqnarray}
&& V_R(p',p'k) =\frac{4 \pi ^2 a}{\epsilon \, m (\pi -2 a \mu )}+\frac{2 \pi ^3 a^2 r_e \text{\textit{k}}^2}{m (\pi -2 a \mu )^2} \nonumber\\
&&  \qquad + \epsilon\left[\frac{\pi ^4 a^3 r_e^2 \text{\textit{k}}^4}{m (\pi -2 a \mu
   )^3}
    +\frac{\pi  a r_e^2 \left(3 (\pi -2 a \mu )
   \text{\textit{k}}^2-2 a \mu ^3\right)}{24 \beta  v_2 (\pi -2 a
   \mu )^2}-\frac{\pi  a r_e^2
   \left(p^2+p'^2\right)}{16 \beta  v_2 (\pi -2 a \mu)}
   \right]+\cdots,
\label{VRKSW}
\end{eqnarray}
where we have
symbolically included here the factors of $\epsilon$ to account for
the orders of various terms. It fulfills the standard KSW counting
with $c_R \sim \epsilon^{-1}$, $c_{ER} \sim \epsilon^{-2}$, $c_{EER}
\sim \epsilon^{-3}$, except for the coupling $c_{2R}  \sim \epsilon^{-1}$ which is
suppressed by one power of the soft scale relative to the KSW
scaling. Exact KSW power counting with  $c_{2R}  \sim \epsilon^{-2}$
is restored for the choice of $\beta \sim \epsilon$.

On the other hand, renormalized couplings of natural size $\sim \epsilon^0$
are obtained for another choice of subtraction points. In fact, taking
\begin{equation}
\mu \sim \Lambda_H , \quad  \mu_{1, \ldots , 5} \sim \epsilon \,,
\end{equation}
and setting $\beta\sim\epsilon^0$, we obtain for the renormalized potential:
\begin{eqnarray}
V_R(p',p,k)=\frac{4 \pi ^2 a}{m(\pi -2 a \mu)}+ \epsilon^2 \left[\frac{\pi ^2 a r_e \text{\textit{k}}^2 \left(16 \pi  a
   \beta  v_2+m r_e\right)}{8 \beta  m v_2 (\pi -2 a \mu
   )^2}-\frac{\pi  a r_e^2
   \left(p^2+p'^2\right)}{16 \beta  v_2 (\pi -2 a \mu
   )}\right] +\cdots.
\label{VRW}
\end{eqnarray}
Equation (\ref{VRW}) corresponds to
renormalized coupling constants of natural size as assumed in Weinberg's power counting.
The scaling of the coupling constants differs from the one suggested
by the standard Wilsonian RG analysis since the latter uses a single
cutoff scale and does not cover
that region of the multi-dimensional space of renormalization scale parameters
which corresponds to Weinberg's power counting. Moreover, for the choice of
the renormalization conditions corresponding to  Weinberg's power counting,
the scaling of renormalized coupling constants is the same regardless of whether
the scattering length is natural or unnaturally large (i.e. the potential provides
a smooth interpolation between the two regimes). Notice further that in case of an
unnaturally large scattering length, Weinberg's power counting does {\it not}
correspond to the expansion of the amplitude around the trivial fixed
point  as it is sometimes erroneously claimed.

It is instructive to look at the perturbative expansion of the
amplitude $T$ of Eq.~(\ref{rentmp2}) within
Weinberg's approach. Taking $\mu\sim\epsilon^0$, $\mu_i\sim\epsilon $
we obtain  for $k\sim \epsilon$
\begin{eqnarray}
T(k) &=& \frac{1}{1/c_{\rm R} + m (2 \mu +i \pi  k)/(4 \pi^2)}
\nonumber\\
&+&  \frac{16 \pi ^2 \left[3 \pi ^2 k^2 c_{\text{ER}}+c_{\rm 2 R} \left(m
   c_{\rm R} \left(3 k^2 \left(\mu -\mu _2\right)-\mu _1^3\right)+6
   \pi ^2 k^2\right)\right]}{3 \left[4 \pi ^2+m c_{\rm R} (2 \mu +i
   \pi  k)\right]{}^2}+\cdots \,.
\label{invamp0pert}
\end{eqnarray}
Naively one might be tempted to conclude that for $\mu$ of the order of the hard scale, the LO amplitude
\begin{equation}
T_{\rm LO} = \frac{1}{1/c_{\rm R} + m (2 \mu +i \pi  k)/(4 \pi^2)}
\end{equation}
is of order $\epsilon^0$. However, due to the cancelation between $\mu$ and $1/c_{\rm R}$,
it is actually of order $\epsilon^{-1}$ while the next term in the expansion of the amplitude
is of the order $\epsilon^0$, etc. Notice that as already pointed out
in section \ref{sec:VTcount}, the appearance of cancelations in the
amplitude beyond naive dimensional analysis is unavoidable in the case
of an unnaturally large scattering length.

In the next section, we extend our analysis to another exactly
solvable toy model with a long-range interaction and show that
all our conclusions remain valid in this case too.

\section{A toy model with a long-range interaction}
\label{sec:long}

\begin{figure}[t]
\begin{center}
\epsfig{file=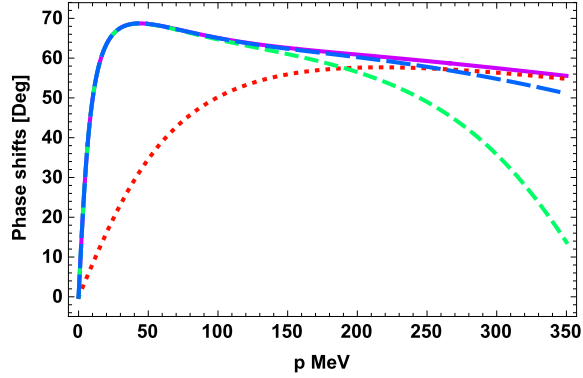,scale=0.6}
\caption{S-wave phase shift of the toy model with a long-range
  interaction  as a function of the momentum in the center-of-mass frame.
%in the spin-singlet channel of the NN interaction.
 The solid (magenta) and the dotted (red) lines correspond to the exact phase shifts
of the toy model and to the switched-off long-range potential, respectively.
The dashed (green) and long-dashed (blue) lines represent phase shifts of the NLO EFT for
the choices of the renormalization scale $\mu=135$ MeV and $\mu=750$ MeV, respectively.}
\label{fig:3}
\end{center}
\end{figure}

Consider a spin-singlet $S$-wave interaction of two nucleons specified by the
following exactly solvable toy-model potential (for more details,
including the form of corresponding LS equation, see 
Ref.~\cite{Epelbaum:2009sd})
\begin{eqnarray}
V(p,p') &=& v_l F_l(p) F_l(p')+v_s F_s(p) F_s(p'),\nonumber\\
F_l(p) &\equiv & \frac{\sqrt{p^2+m_s^2}}{p^2+m_l^2}, \ \ \ F_s(p) \equiv \frac{1}{\sqrt{p^2+m_s^2}},
\label{potSLR2}
\end{eqnarray}
where $m_l$ and $m_s$ are the small and large mass scales corresponding to the long- and
the short-range interaction.  We choose the strength of the long-range interaction
$v_l=\alpha\, m_l^4$  such that the LO long-range potential is of order zero
when $m_l$ and the momenta are counted as soft quantities.
To generate the phase shifts similar to those of the $^1S_0$ partial wave of $np$
scattering, we use $m_l=135$ MeV, $m_s=750$ MeV and tune the coupling constants  $v_s$
and $\alpha$ such that the scattering length and the effective range are
$a=-1/(8 \ {\rm MeV})$ and $r_e=1/(100 \ {\rm MeV})$, respectively. For the
switched-off long-range interaction the ERE parameters turn out to be
$a=-1/(68.71 \ {\rm MeV})$ and $r_e=1/(343.53 \ {\rm MeV})$. The resulting phase shifts
are shown in Fig.~\ref{fig:3}.

To reproduce the phase shifts of the ``underlying theory'' in the EFT approach we count
$m_l$ and the three-momenta as small quantities and consider the following NLO effective
potential
\begin{eqnarray}
V_{\rm EFT}(p,p') & = & c  +  \frac{\alpha_0\, m_l^4}{\left(m_l^2+p^2\right)
   \left(m_l^2+p'^2\right)} \nonumber\\
   &+& c_2(p^2+p'^2)+d \, m_l^2  +\frac{\alpha'_0  m_l^4 \left(p^2+p'^2\right)+\alpha''_0 m_l^6}{2 m_s^2
   \left(m_l^2+p^2\right) \left(m_l^2+p'^2\right)},
\label{EFTpot}
 \end{eqnarray}
 where $\alpha_0$, $\alpha'_0$ and $\alpha''_0$ are bare parameters of the long-range
part and $c$, $c_2$ and $d$ are the bare couplings of the contact interaction
(short-range) terms.
All divergences appearing in the NLO amplitude can be absorbed in the renormalization
of the parameters of the potential of Eq.~(\ref{EFTpot}) provided that the NLO terms
of the potential are treated perturbatively.  Notice that due to the separability of
the long-range part of the considered toy model potential, the parameters of the
long-range part also need to be renormalized in order to remove divergences of
loop diagrams  obtained by iterations of the potential.
We treat the full NLO potential non-perturbatively by substituting it into
the integral equation and applying
the subtractive renormalization.  To match the
   results for the scattering amplitude with the ones based on the
   underlying model,
we take the renormalized couplings of the long-range EFT
potential as $\alpha_R=\alpha'_R=\alpha''_R=\alpha$.
 In exact analogy to the above
considered case of the contact interactions alone, we discriminate between the
renormalization scale $\mu$, corresponding to the LO interaction, and all other
renormalization scales which we put equal to zero. Tuning the
renormalized couplings of the contact interaction terms  $c_{\rm R}$ and
$c_{\rm 2R}$ to reproduce the scattering length and the
effective range (we take $d_R=0$ as it cannot be disentangled from
$c_{\rm R}$) we obtain the values
\begin{eqnarray}
c_{\rm R} &=& -3.770 \, \frac{4\pi}{m \, m_s}~, \qquad c_{\rm 2 R}=
              7.727\, \frac{4\pi}{m \, m_s^3} \qquad {\rm for}~~~  \mu=135 \ {\rm MeV},\nonumber\\[4pt]
c_{\rm R} &=& - 0.916 \, \frac{4\pi}{m \, m_s}~, \qquad c_{\rm 2 R}=
              0.447 \, \frac{4\pi}{m \, m_s^3} \qquad
{\rm for}~~~   \mu=750 \ {\rm MeV},
\end{eqnarray}
where the factors
of $4\pi/m$ emerge from the employed normalization of the potential and
T-matrix, see Eq. (\ref{invamp0}).
Thus, by choosing the renormalization scale of the order of the soft scale $m_l$ of the
problem, the renormalized couplings of the contact interactions are enhanced,
as suggested by KSW-like counting (and by the standard Wilsonian RG
analysis with a single cutoff scale), and they are natural for the
renormalization scale of the order of the hard scale.

We also plot in
Fig.~\ref{fig:3} the resulting phase shifts at LO and NLO for
$\mu=m_l=135 $ MeV and $\mu=m_s=750$ MeV. Notice that the dependence
of the phase shifts on the renormalization scale $\mu$ emerges as a
consequence of the non-perturbative inclusion of the subleading contributions
to the potential. One observes that the
choice of the renormalization scale of the order of the hard scale in
the problem leads to a better reproduction of the phase shift at
higher energy, fully in line with the findings of Ref.~\cite{Epelbaum:2015sha}. It is
shown in that paper (in the framework of pionless EFT), that the
choice of the subtraction scale $\mu$ of the order of the soft scale
results in enhanced scheme- and $\mu$-dependent contributions to
the phase shifts for the case of an unnaturally large scattering
length.

\begin{figure}[t]
\begin{center}
\epsfig{file=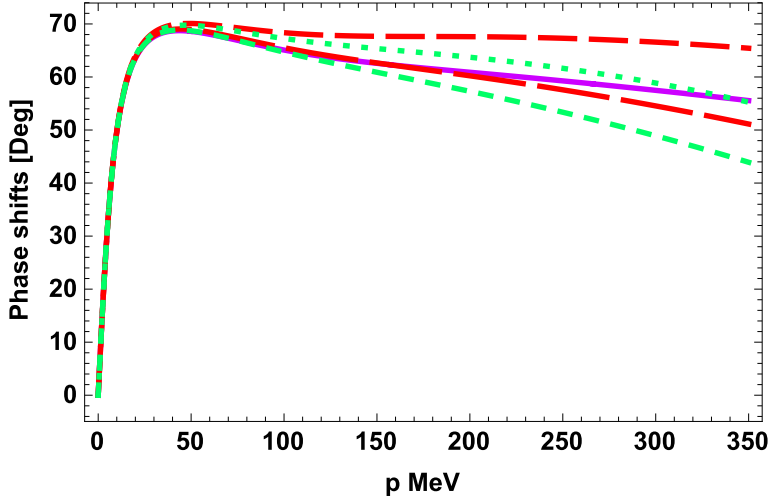,scale=0.8}
\caption{
S-wave phase shift of the toy model with a long-range
  interaction  as a function of the  momentum in the center-of-mass frame.
%Phase shifts in the S-wave spin-singlet channel of $np$ scattering.
The solid (magenta) line corresponds to the exact phase shifts.
The dotted and short-dashed (green) lines represent the LO and NLO phase shifts
for the choice of the cutoff $\Lambda=500$~MeV, respectively, and
the middle- and long-dashed (red) lines  represent analogous results for $\Lambda=800$~MeV.
}
\label{fig:5}
\end{center}
\end{figure}

While the considered model with a long-range interaction of a
separable type leads to simple, analytical expressions for the
scattering amplitude, which can be explicitly renormalized by replacing
the bare coupling constants in terms of the renormalized ones, the
situation is much more complicated for a realistic case of
the two-nucleon force, whose long-range tail is governed by the
OPE. It is not feasible to perform subtractive
renormalization of all iterations of the LS equation when the OPE is
treated non-perturbatively. Instead, one usually introduces a finite
cutoff chosen of the order of the hard scale in the problem and
performs \emph{implicit} renormalization by (effectively) expressing the bare
coupling constants in terms of observable quantities such as NN phase
shifts. We have followed the same cutoff EFT approach for our toy
model and calculated the LO and NLO phase
shifts. Our results are displayed in Fig.~\ref{fig:5} for the cutoffs $\Lambda=500$ MeV and
$\Lambda=800$ MeV  along with the phase shifts of the underlying
model. The quality of the description of the phase shifts at NLO is
comparable to the ones obtained after explicit renormalization of the
amplitude as visualized in Fig.~\ref{fig:3}.

Finally, we have analyzed the exact Wilsonian RG trajectory of the cutoff regularized
potential corresponding to the exact potential of Eq.~(\ref{potSLR2}). By applying
the sharp cutoff to the  LS  equation we obtained the
corresponding cutoff dependent potential leading to the cutoff independent
off-shell scattering amplitude. This energy-dependent potential  satisfies the
RG equation of Eq.~(\ref{RGE1}). In exact analogy to the case of the contact
interactions alone, the Wilsonian RG analysis leads to the KSW-like power counting
for couplings of the contact interactions also in the presence of
a long-range interaction.
These results are in line with the ones found earlier by Birse
and collaborators~\cite{Birse:1998dk}.
However, we stress again that  the Wilsonian RG analysis of
Ref.~\cite{Birse:1998dk} does not cover the full  range in the space of
renormalization  scale parameters, in particular, the range corresponding to Weinberg's power counting.
This means that both the KSW and the Weinberg approach are consistent with the
exact RG but correspond to different choices of the subtraction scales.

\section{Summary and conclusions}
\label{sec:summ}

Using two examples of exactly renormalizable toy model nucleon-nucleon potentials
we have compared the most general subtractive renormalization and the Wilsonian
renormalization group (RG) approach of Ref.~\cite{Birse:1998dk}. We find that the
scaling of coupling constants uncovered by the Wilsonian RG analysis corresponds
to the choice of the renormalization scheme when all subtraction points are
chosen of the order of the soft scale of the problem. This scaling is also shared by
the KSW power counting of Ref.~\cite{Kaplan:1998tg}. On the other hand,
by choosing the renormalization point corresponding to the coupling constant
of the momentum- and energy-independent contact interaction of the order
of the hard scale of the problem while taking all other renormalization points
of the order of the soft scale, one recovers  Weinberg's power
counting~\cite{Weinberg:1990rz,Weinberg:1991um} with renormalized coupling constants being
of natural size both for natural as well as unnaturally
large scattering lengths.

In the KSW approach of Ref.~\cite{Kaplan:1998tg}, dimensional
regularization along with PDS subtraction scheme
has been used. Therefore, within this approach, all renormalization points are
taken either zero or  of the order of the scale of dimensional regularization.
In standard Wilsonian RG approach one also uses a single scale, the
cutoff parameter, and hence in both these cases one studies the behaviour of
couplings  in a one-parameter subspace of the multi-dimensional space of the
renormalization group of the corresponding EFT. As a result of this restriction,
in both the KSW and the standard Wilsonian RG approaches one does not
cover that area in the space of renormalization parameters which is appropriate
for the Weinberg approach to nucleon-nucleon scattering problem for the case of
an unnaturally large scattering length.
We also emphasize that in the case of an unnaturally large scattering length, Weinberg's
power counting does {\it not} correspond to the expansion around the trivial fixed
point as it is sometimes claimed.

When performing realistic chiral EFT calculations of NN scattering,
the Lippmann-Schwinger equation is usually regularized with a finite
cutoff, chosen of the order of the hard scale in the problem
as done e.g.~in Refs.~\cite{Epelbaum:2014efa,Epelbaum:2014sza}.
It is not known to us how to practically implement  a subtractive renormalization
with the pion-exchange potentials being treated non-perturbatively. In such
calculations, renormalization is carried out \emph{implicitly} by adjusting
the bare low-energy constants to experimental data or phase shifts,
see Ref.~\cite{Epelbaum:2015pfa} for a discussion. We conjecture that this approach is
equivalent to the choice of renormalization conditions specified
above, i.e.~with the subtraction scale corresponding to the LO (higher-order) contact
interactions chosen of the order of the hard (soft) scales of the problem,
provided the determined bare LECs are of a natural
size. This conjecture can be easily verified for the case of the pionless EFT
approach, where the analytical expressions for the scattering
amplitude are available.

\acknowledgments

EE and JG  are indebted to M.~Birse for numerous illuminating discussions on the Wilsonian RG
approach. EE and JG  thank the [Department of Energy's] Institute for Nuclear Theory at the University of
Washington for its hospitality and the Department of Energy for partial support during the completion
of this work.
This work was supported in part by BMBF (contract No.~05P2015 - NUSTAR
R\&D), by DFG and NSFC through funds provided to the
Sino-German CRC 110 ``Symmetries and the Emergence of Structure in QCD" (NSFC
Grant No.~11621131001, DFG Grant No.~TRR110), by the Georgian Shota Rustaveli National
Science Foundation (grant FR/417/6-100/14) and by the CAS President's International
Fellowship Initiative (PIFI) (Grant No.~2017VMA0025).

\appendix
\section{Various expressions}

The potential of section~\ref{sec:pionless} in the new parametrization:
\begin{eqnarray}
V\left(p',p,k\right) & = &  \frac{P_1}{P_2}\,,\nonumber\\
P_1 & = & - c_2^2 k m \left(c_2^2-c c_{\text{pp}}\right)
   \left(k^2-p^2\right) \left(k^2-p'^2\right) \ln
   \frac{\Lambda -k}{k+\Lambda } \nonumber\\
&-& 4 \pi ^2
   \biggl[c_2^2 \left(k^2 c_{\text{E}}+k^4 c_{\text{EE}}-k^2
   c_{\text{Ep}} p'^2-k^2 p^2 c_{\text{Ep}}+p^2
   c_{\text{pp}} p'^2+c\right) \nonumber\\
&+& k^2 c_{\text{pp}}
   \left(k^2 c_{\text{E}}^2+c c_{\text{E}}-c k^2
   c_{\text{EE}}\right) +c_2 k^2 \left(c_{\text{E}}
   c_{\text{pp}} \left(p'^2+p^2\right)-2
   c_{\text{Ep}} \left(k^2 c_{\text{E}}+c\right)\right)\nonumber\\
&+& c k^4
   c_{\text{Ep}}^2+c_2^3
   \left(p'^2+p^2\right)\biggr]\,,\nonumber\\
P_2 & = & k m \ln \frac{\Lambda -k}{k+\Lambda } \biggl\{k^2
   \biggl[c_2^2 \left(c_{\text{E}}+k^2 \left(c_{\text{EE}}-2
   c_{\text{Ep}}+c_{\text{pp}}\right)\right)+2 c_2 k^2
   c_{\text{E}} \left(c_{\text{pp}}-c_{\text{Ep}}\right) \nonumber\\
&+& k^2
   c_{\text{E}}^2 c_{\text{pp}}+2 c_2^3\biggr] +c \left(k^2
   c_{\text{pp}} \left(c_{\text{E}}-k^2
   c_{\text{EE}}\right)+k^4 c_{\text{Ep}}^2-2 c_2 k^2
   c_{\text{Ep}}+c_2^2\right)\biggr\} \nonumber\\
&-& 4 \pi ^2 \left(k^2
   c_{\text{pp}} \left(c_{\text{E}}-k^2
   c_{\text{EE}}\right)+k^4 c_{\text{Ep}}^2-2 c_2 k^2
   c_{\text{Ep}}+c_2^2\right)\,.
\end{eqnarray}

The on-shell amplitude in the new parametrization:
\begin{eqnarray}
\frac{1}{T} & = & \frac{N_0}{D_0} +  \frac{i  k\,m}{4 \pi } \,,\nonumber\\
N_0 & = & 90 \pi ^2 m \Lambda  \biggl\{k^4 \biggl[c_2 \left(c_2
   c_{\text{EE}}-2 \left(c_{\text{E}}+c_2\right)
   c_{\text{Ep}}\right)+\left(c_{\text{E}}+c_2\right){}^2
   c_{\text{pp}}\biggr]\nonumber\\
&+& c \biggl[k^2 c_{\text{E}}
   c_{\text{pp}}+k^4 \left(c_{\text{Ep}}^2-c_{\text{EE}}
   c_{\text{pp}}\right)-2 c_2 k^2
   c_{\text{Ep}}+c_2^2\biggr] +c_2^2 k^2 \left(c_{\text{E}}+2
   c_2\right)\biggr\}\nonumber\\
&+& 180 \pi ^4 \biggl[k^2 c_{\text{E}}
   c_{\text{pp}}+k^4 \left(c_{\text{Ep}}^2-c_{\text{EE}}
   c_{\text{pp}}\right)-2 c_2 k^2 c_{\text{Ep}}+c_2^2\biggr] \nonumber\\
&+& 30
   \pi ^2 c_2 m \Lambda ^3 \biggl[k^2 \left(2 c_{\text{E}}
   c_{\text{pp}}+c_2 \left(c_{\text{pp}}-2
   c_{\text{Ep}}\right)\right)+2 c_2^2\biggr]\nonumber\\
&+& 15 c_2^2 k^2 m^2
   \Lambda ^4 \left(c_2^2-c c_{\text{pp}}\right)-4 c_2^2 m^2
   \Lambda ^6 \left(c_2^2-c c_{\text{pp}}\right)+18 \pi ^2 c_2^2
   m \Lambda ^5 c_{\text{pp}}\,,\nonumber\\
D_0 & = &  6 \pi ^2 \biggl\{30 \pi ^2 \biggl[k^4 \left(c_2 \left(c_2
   c_{\text{EE}}-2 \left(c_{\text{E}}+c_2\right)
   c_{\text{Ep}}\right)+\left(c_{\text{E}}+c_2\right){}^2
   c_{\text{pp}}\right)  \nonumber\\
&+& c \left(k^2 c_{\text{E}}
   c_{\text{pp}}+k^4 \left(c_{\text{Ep}}^2-c_{\text{EE}}
   c_{\text{pp}}\right)-2 c_2 k^2
   c_{\text{Ep}}+c_2^2\right)+c_2^2 k^2 \left(c_{\text{E}}+2
   c_2\right)\biggr]\nonumber\\
&+& 5 c_2^2 k^2 m \Lambda ^3 \left(c_2^2-c
   c_{\text{pp}}\right)-3 c_2^2 m \Lambda ^5 \left(c_2^2-c
   c_{\text{pp}}\right)\biggr\}\,.
 \label{2}
\end{eqnarray}

The bare couplings of the new parametrization expressed in terms of
the renormalized ones:
\begin{eqnarray}
x&:=& \frac{2 \pi
   ^2}{c_{\rm 2R}^2-c_{\rm ppR} c_{\rm R}}, \nonumber\\
c & = &  \frac{18 \pi ^2 \left[ m \left(\mu_3^5-\Lambda
^5\right)-5 c_{\rm R} x \right]}{9
   \left[ m (\mu -\Lambda ) - c_{\rm ppR} x
   \right]
   \left[ m \left(\mu_3^5-\Lambda
   ^5\right) -5 c_{\rm R} x\right]
   -5 \left[
   3 c_{\rm 2R} x +
   m \left(\mu_1^3-\Lambda ^3\right)\right]^2}\,,\nonumber\\
c_2 & = & \frac{30 \pi ^2 \left[ 3
   c_{\rm 2R} x +
   m \left(\mu_1^3-\Lambda ^3\right)\right]}{5
   \left(3 c_{\rm 2R} x + m \left(\mu_1^3-\Lambda
   ^3\right)\right)^2-9 \left( m (\mu -\Lambda
   ) - c_{\rm ppR} x \right) \left( m \left(\mu_3^5-\Lambda ^5\right)-5 c_{\rm R} x \right)} \,,\nonumber\\
c_{\rm pp} & = &  \frac{90 \pi ^2 \left( m (\mu -\Lambda
   ) -  c_{\rm ppR} x \right)}{9 \left( m (\mu -\Lambda
   )- c_{\rm ppR} x \right) \left( m \left(\mu_3^5-\Lambda ^5\right)-5 c_{\rm R} x\right)-5
   \left(3 c_{\rm 2R} x + m \left(\mu_1^3-\Lambda
   ^3\right)\right)^2} \,,\nonumber\\
c_{E} & = & \frac{-1}{\left[5 \left(3
   c_{\rm 2R} x +
   m \left(\mu_1^3-\Lambda ^3\right)\right)^2-9
   \left( m (\mu -\Lambda )- c_{\rm ppR} x
   \right)
   \left( m \left(\mu_3^5-\Lambda
   ^5\right)-5 c_{\rm R} x \right)\right]^2}\nonumber\\
& \times &  \Biggl\{30 \pi ^2 \left( 3
   c_{\rm 2R} x +
   m \left(\mu_1^3-\Lambda ^3\right)\right)
   \Biggr[5 \left(3
   c_{\rm 2R} x +
   m \left(\mu_1^3-\Lambda ^3\right)\right)\nonumber\\
& \times &
   \left( m \left(\mu_4^3-\Lambda
   ^3\right)-\frac{3 \left(c_{ER} c_{\rm 2R}^2-2 c_{EpR}
   c_{\rm R} c_{\rm 2R}+c_{ER} c_{\rm ppR} c_{\rm R}\right)
   x}{c_{\rm 2R}^2}\right) \nonumber\\
& - & 18 \left( m (\mu_2-\Lambda )-\frac{(c_{\rm 2R}
c_{EpR}-c_{ER}
   c_{\rm ppR}) x}{c_{\rm 2R}}\right) \left( m \left(\mu_3^5-\Lambda ^5\right)-5 c_{\rm R} x \right)\Biggr]
\Biggr\}
\,,\nonumber\\
c_{\rm EE} & = & \frac{90 \pi ^2}{\left\{
5 \left[3
   c_{\rm 2R} x+
   m \left(\mu_1^3-\Lambda ^3\right)\right]^2-
   9 \left[ m (\mu -\Lambda )-c_{\rm ppR} x\right]
   \left[ m \left(\mu_3^5-\Lambda
   ^5\right)-5 c_{\rm R} x \right]
   \right\}^3}\nonumber\\
& \times & \Biggl\{25 \left[\frac{\left(c_{EER}
   c_{\rm 2R}^2-2 c_{EpR} c_{ER} c_{\rm 2R}+c_{ER}^2
   c_{\rm ppR}+\left(c_{EpR}^2-c_{EER}
   c_{\rm ppR}\right)
   c_{\rm R}\right) x }{c_{\rm 2R}^2}+ m (\Lambda -\mu_5)\right]\nonumber\\
& \times & \left[ 3
   c_{\rm 2R} x+
   m \left(\mu_1^3-\Lambda
   ^3\right)\right]^4
   +50 \left[ m (\mu_2-\Lambda )-\frac{ (c_{\rm 2R} c_{EpR}-c_{ER}
   c_{\rm ppR}) x}{c_{\rm 2R}}\right]\nonumber\\
& \times & \left[ m \left(\mu_4^3-\Lambda ^3\right)-\frac{3
\left(c_{ER}
   c_{\rm 2R}^2-2 c_{EpR} c_{\rm R} c_{\rm 2R}+c_{ER}
   c_{\rm ppR} c_{\rm R}\right) x
   }{c_{\rm 2R}^2}\right]
   \left[3 c_{\rm 2R} x + m \left(\mu_1^3-\Lambda
   ^3\right)\right]^3 \nonumber\\
& - & 5 \Biggl[ 27 \left( m
   \left(\mu_3^5-\Lambda ^5\right)- 5 c_{\rm R} x \right)
   \left( m (\mu_2-\Lambda )-\frac{
   (c_{\rm 2R} c_{EpR}-c_{ER} c_{\rm ppR}) x
   }{c_{\rm 2R}}\right)^2 \nonumber\\
& + & 5 \left( m (\mu -\Lambda )- c_{\rm ppR}
   x \right)
   \left( m \left(\mu_4^3-\Lambda
   ^3\right)-\frac{6 \left(c_{ER} c_{\rm 2R}^2-2 c_{EpR}
   c_{\rm R} c_{\rm 2R}+c_{ER} c_{\rm ppR} c_{\rm R}\right)
   \pi ^2}{c_{\rm 2R}^4-c_{\rm 2R}^2 c_{\rm ppR}
   c_{\rm R}}\right)^2 \nonumber\\
& - & 9 \left[ m (\mu -\Lambda
   ) - c_{\rm ppR} x \right] \left[ m \left(\mu_3^5-\Lambda ^5\right)
   -5 c_{\rm R} x \right]\nonumber\\
& \times &
   \left( m (\mu_5-\Lambda )-\frac{
   \left(c_{\rm EER} c_{\rm 2R}^2-2 c_{\rm EpR} c_{\rm ER}
   c_{\rm 2R}+c_{\rm ER}^2
   c_{\rm ppR}+\left(c_{\rm EpR}^2-c_{\rm EER}
   c_{\rm ppR}\right) c_{\rm R}\right) x
   }{c_{\rm 2R}^2 }\right)
   \Biggr]
   \nonumber\\
& \times &  \left( 3
   c_{\rm 2R} x +
   m \left(\mu_1^3-\Lambda
   ^3\right)\right)^2+90 \left( m (\mu -\Lambda
   ) - c_{\rm ppR} x \right)\nonumber\\
& \times &  \left( m (\mu_2-\Lambda )-\frac{(c_{\rm 2R}
c_{\rm EpR}-c_{\rm ER}
   c_{\rm ppR}) x}{c_{\rm 2R}}\right) \left( m \left(\mu_3^5-\Lambda ^5\right) - 5 c_{\rm R} x \right) \nonumber\\
& \times &
   \left[ m \left(\mu_4^3-\Lambda
   ^3\right)-\frac{3 \left(c_{\rm ER} c_{\rm 2R}^2-2 c_{\rm EpR}
   c_{\rm R} c_{\rm 2R}+c_{ER} c_{\rm ppR} c_{\rm R}\right)
   x }{c_{\rm 2R}^2}\right]
   \left[ 3
   c_{\rm 2R} x  +
   m \left(\mu_1^3-\Lambda ^3\right)\right] \nonumber\\
& + & 81
   \left[ c_{\rm ppR} x   + m (\Lambda -\mu )\right]
   \left[
   \frac{(c_{\rm 2R} c_{\rm EpR}-c_{\rm ER}
   c_{\rm ppR}) x }{c_{\rm 2R}}+ m (\Lambda -\mu_2)
   \right]^2
   \nonumber\\
& \times & \left[
m \left(\mu_3^5-\Lambda ^5\right) - 5
c_{\rm R} x
   \right]^2\Biggr\}\,,\nonumber\\
c_{Ep} & = & \frac{90 \pi ^2}{
\left\{5 \left[
3 c_{\rm 2R} x +
   m \left(\mu_1^3-\Lambda ^3\right)\right]^2-9
   \left[ m (\mu -\Lambda ) - c_{\rm ppR} x \right]
   \left[ m \left(\mu_3^5-\Lambda
   ^5\right) - 5 c_{\rm R} x \right]
   \right\}^2}
   \nonumber\\
& \times &  \Biggl\{5 \left[
\frac{(c_{\rm 2R}
   c_{\rm EpR}-c_{\rm ER} c_{\rm ppR}) x}{c_{\rm 2R}}+ m (\Lambda
   -\mu_2)\right]
   \left[ m
   \left(\Lambda ^3-\mu_1^3\right) - 3 c_{\rm 2R} x \right]^2 \nonumber\\
& - & 9 \left[ m (\mu -\Lambda ) - c_{\rm ppR} x \right]
   \left[ m (\mu_2-\Lambda )-\frac{
   (c_{\rm 2R} c_{\rm EpR}-c_{\rm ER} c_{\rm ppR}) x
   }{c_{\rm 2R}}\right]
   \nonumber\\
& \times & \left[ m \left(\mu_3^5-\Lambda
   ^5\right) - 5 c_{\rm R} x \right]
   + 5 \left[ m (\mu -\Lambda
   ) - c_{\rm ppR} x\right]   \label{csbare} \\
& \times &  \left[ 3
   c_{\rm 2R} x +
   m \left(\mu_1^3-\Lambda ^3\right)\right]
   \Biggl[ m \left(\mu_4^3-\Lambda
   ^3\right) -  \frac{3 \left(c_{\rm ER} c_{\rm 2R}^2-2 c_{\rm EpR}
   c_{\rm R} c_{\rm 2R}+c_{\rm ER} c_{\rm ppR} c_{\rm R}\right)
   x }{c_{\rm 2R}^2}\Biggr]\Biggr\} \,.
\nonumber
\end{eqnarray}

The renormalization scheme dependence of the renormalized
couplings has the form
\begin{eqnarray} c_{\rm R} & = &  \frac{C_{\rm 4R}(\mu_3)}{C_{\rm 4R}(\mu_3)
   C_{\rm R}(\mu )-C_{\rm 2R}(\mu_1)^2}\,,\nonumber\\
c_{\rm 2R} & = &  \frac{C_{\rm 2R}(\mu_1)}{C_{\rm 2R}(\mu_1)^2-C_{\rm 4R}(\mu_3) C_{\rm R}(\mu )}\,,\nonumber\\
c_{\rm ppR} & = &  \frac{C_{\rm R}(\mu )}{C_{\rm 4R}(\mu_3)
   C_{\rm R}(\mu )-C_{\rm 2R}(\mu_1)^2}\,,\nonumber\\
c_{\rm ER} & = & \frac{C_{\rm 2R}(\mu_1) \left[ 2 C_{\rm 2ER}(\mu_2)
C_{\rm 4R}(\mu_3)-C_{\rm 2R}(\mu_1)
   C_{\rm 4ER}(\mu_4)\right]
   }{\left[ C_{\rm 2R}(\mu_1)^2-C_{\rm 4R}(\mu_3) C_{\rm R}(\mu
   )\right]^2}\,,\nonumber\\
c_{\rm EER} & = & -\frac{1}
   {\left[c_{\rm 2R}(\mu_1)^2-C_{\rm 4R}(\mu_3) C_{\rm R}(\mu
   )\right]^3} \,\Biggl\{ C_{\rm 4EER}(\mu_5) C_{\rm 2R} (\mu_1)^4 \nonumber\\
& - & 2 C_{\rm 2ER}(\mu_2)
   C_{\rm 4ER}(\mu_4) c_{\rm 2R}(\mu_1)^3+\biggl[3 C_{\rm 4R}(\mu_3)
   C_{\rm 2ER}(\mu_2)^2 \nonumber\\
& + & \left(C_{\rm 4ER}(\mu_4)^2
   -C_{\rm 4EER}(\mu_5) C_{\rm 4R}(\mu_3)\right) C_{\rm R}(\mu )\biggr] c_{\rm 2R}(\mu_1)^2 \nonumber\\
& - & 2 C_{\rm 2ER}(\mu_2)
   C_{\rm 4ER}(\mu_4) C_{\rm 4R}(\mu_3)
   C_{\rm R}(\mu ) C_{\rm 2R}(\mu_1)
 + C_{\rm 2ER}(\mu_2)^2 C_{\rm 4R}(\mu_3)^2 C_{\rm R}(\mu )\Biggr\}\,,\nonumber\\
c_{\rm EpR} & = &  \frac{C_{\rm 2R}(\mu_1) C_{\rm 4ER}(\mu_4)
   C_{\rm R}(\mu )-C_{\rm 2ER}(\mu_2)
   \left[C_{\rm 2R}(\mu_1)^2+C_{\rm 4R}(\mu_3) C_{\rm R}(\mu )\right]}{\left[C_{\rm 2R}(\mu_1)^2-C_{\rm 4R}(\mu_3) C_{\rm R}(\mu
   )\right]^2}\,.
   \label{runningcouplings}
\end{eqnarray}

The renormalized couplings corresponding to the parameters of Eq.~(\ref{parameters})
\begin{eqnarray}
c_{\rm R} &=& \frac{n_1}{d_1},\nonumber\\
n_1 &=& {576 \pi ^2 a v_2 \left(m^2 \mu _3^5 r_e^2-640 \pi ^3 v_2
   \beta ^2\right)},\nonumber\\
d_1 &=& 9 \pi  a m^3 \mu _3^5 r_e^4+16 m v_2 \biggl[m
   r_e^2 \left(10 a m \mu _1^6+9 m \mu _3^5 (\pi -2 a \mu )-120
   \pi ^2 a \beta  \mu _1^3\right) \nonumber\\
&-&5760 \pi ^3 v_2 \beta ^2 (\pi
   -2 a \mu )\biggr],\nonumber\\[6pt]
c_{\rm 2R} &=& \frac{n_2}{d_2},\nonumber\\
n_2 &=&{960 \pi ^2 a v_2 r_e^2 \left(6 \pi ^2 \beta -m \mu
   _1^3\right)},\nonumber\\
d_2 &=&  9 \pi  a m^2 \mu _3^5 r_e^4+16 m v_2 r_e^2
   \left[10 a m \mu _1^6+9 m \mu _3^5 (\pi -2 a \mu )-120 \pi ^2
   a \beta  \mu _1^3\right]\nonumber\\
& &-92160 \pi ^3 v_2^2 \beta ^2 (\pi -2
   a \mu ),\nonumber\\[6pt]
c_{ER} &=& \frac{n_3}{d_3},\nonumber\\
n_3 &=&1920 \pi ^2 a v_2 r_e \left(m \mu _1^3-6 \pi ^2 \beta
   \right) \biggl\{-960 \pi ^2 v_2 \beta  \biggl[a m^2 \mu _1^3
   r_e^3 \nonumber\\
&+&96 \pi  m v_2 \beta  \left(\pi -2 a \mu _2\right)
   r_e+1536 \pi ^3 a v_2^2 \beta ^2\biggr] +9 m^2 \mu _3^5 r_e^2
   \biggl[16 m v_2 \left(\pi -2 a \mu _2\right) r_e \nonumber\\
&+&\pi  a m
   r_e^3+256 \pi ^2 a v_2^2 \beta \biggr] +80 a m^2 v_2 \mu _4^3
   r_e^3 \left(m \mu _1^3-6 \pi ^2 \beta \right)\biggr\},\nonumber\\
d_3 &=& m
   \biggl\{9 \pi  a m^2 \mu _3^5 r_e^4  +16 m v_2 r_e^2 \biggl[10 a m
   \mu _1^6+9 m \mu _3^5 (\pi -2 a \mu )-120 \pi ^2 a \beta  \mu
   _1^3\biggr] \nonumber\\
&  -&92160 \pi ^3 v_2^2 \beta ^2 (\pi -2 a \mu
   )\biggr\}^2,\nonumber\\[6pt]
c_{\rm ppR} &=& \frac{n_4}{d_4},\nonumber\\
n_4 &=&{90 \pi ^2 m r_e^2 \left(\pi  a r_e^2+16 v_2 (\pi -2 a \mu
   )\right)},\nonumber\\
d_4 &=& -9 \pi  a m^2 \mu _3^5 r_e^4+16 m v_2 r_e^2
   \left(-10 a m \mu _1^6-9 m \mu _3^5 (\pi -2 a \mu )+120 \pi
   ^2 a \beta  \mu _1^3\right) \nonumber \\
  &+&92160 \pi ^3 v_2^2 \beta ^2 (\pi
   -2 a \mu ),\nonumber\\[6pt]
c_{EpR} &=& \frac{n_5}{d_5},\nonumber\\
n_5 & = & 90 \pi ^2 r_e \biggl\{ 9 \pi ^2 a^2 m^3 \mu _3^5 r_e^7+2304
   \pi ^3 a^2 m^2 v_2^2 \beta  \mu _3^5 r_e^4-256 m v_2^2 r_e^3
   \biggl[10 a m^2 \mu _1^6 \left(\pi -2 a \mu _2\right)\nonumber\\
&-& 9 m^2
   \mu _3^5 (\pi -2 a \mu ) \left(\pi -2 a \mu _2\right)+10 a m
   \mu _1^3 \left(24 \pi ^2 a \beta  \left(\mu _2-\mu \right)-m
   \mu _4^3 (\pi -2 a \mu )\right)\nonumber\\
&+& 60 \pi ^2 a m \beta  \mu _4^3
   (\pi -2 a \mu )+360 \pi ^4 a \beta ^2 \left(2 a \mu -4 a \mu
   _2+\pi \right)\biggr] \nonumber\\
&-& 4096 \pi ^2 a v_2^3 \beta  r_e^2
   \biggl[10 a \left(m^2 \mu _1^6-12 \pi ^2 m \beta  \mu _1^3+72
   \pi ^4 \beta ^2\right)-9 m^2 \mu _3^5 (\pi -2 a \mu
   )\biggr]\nonumber\\
&-& 32 \pi  a m^2 v_2 r_e^5 \left(5 a \mu _4^3 \left(6
   \pi ^2 \beta -m \mu _1^3\right)+5 a m \mu _1^6+9 m \mu _3^5
   \left(a \mu +a \mu _2-\pi \right)\right)\nonumber\\
&-& 1474560 \pi ^3 m
   v_2^3 \beta ^2 (\pi -2 a \mu ) \left(\pi -2 a \mu _2\right)
   r_e-23592960 \pi ^5 a v_2^4 \beta ^3 (\pi -2 a \mu
   )\biggr\},\nonumber\\
d_5 & = & \biggl[-9 \pi  a m^2 \mu _3^5 r_e^4+16 m v_2 r_e^2
   \left(-10 a m \mu _1^6-9 m \mu _3^5 (\pi -2 a \mu )+120 \pi
   ^2 a \beta  \mu _1^3\right)\nonumber\\
  &+& 92160 \pi ^3 v_2^2 \beta ^2 (\pi
   -2 a \mu )\biggr]{}^2,\nonumber\\[6pt]
c_{EER} &=& \frac{n_6}{d_6},\nonumber\\
n_6 &= & -90 \pi ^2 \biggl\{81 a^3 m^6 \pi ^3 \mu _3^{10}
   r_e^{12}-144 a^2 m^5 \pi ^2 v_2 \mu _3^5 \biggl[20 a m \mu
   _1^6+9 m \left(2 a \mu +4 a \mu _2-3 \pi \right) \mu _3^5\nonumber\\
&+& 20
   a \left(6 \pi ^2 \beta -m \mu _1^3\right) \mu _4^3\biggr]
   r_e^{10}+41472 a^3 m^5 \pi ^4 \beta  v_2^2 \mu _3^{10}
   r_e^9 \nonumber\\
  &+&256 a m^4 \pi  v_2^2 \biggl[100 a^2 m^2 \mu _1^{12}-200
   a^2 m^2 \mu _4^3 \mu _1^9 \nonumber\\
&-& 20 a m \biggl(9 m \mu _3^5 \left(a
   \mu -6 a \mu _2+a \mu _5+2 \pi \right)-5 a \mu _4^3 \left(m
   \mu _4^3+12 \pi ^2 \beta \right)\biggr) \mu _1^6 \nonumber\\
&-& 120 a m
   \biggl(10 a \pi ^2 \beta  \mu _4^6+3 \mu _3^5 \left(m (a \mu
   -\pi ) \mu _4^3+a \mu _2 \left(m \mu _4^3+24 \pi ^2 \beta
   \right)-6 a \pi ^2 \beta  \left(3 \mu +\mu
   _5\right)\right)\biggr) \mu _1^3\nonumber\\
&+& 9 \biggl(9 m^2 \left(\pi -2 a
   \mu _2\right) \left(-4 a \mu -2 a \mu _2+3 \pi \right) \mu
   _3^{10}\nonumber\\
  &+&240 a \pi ^2 \beta  \biggl(m (a \mu -\pi ) \mu _4^3+a
   \mu _2 \left(m \mu _4^3+18 \pi ^2 \beta \right)\nonumber\\
&-& 3 \pi ^2
   \beta  \left(3 a \mu +a \mu _5+\pi \right)\biggr) \mu
   _3^5+400 a^2 \pi ^4 \beta ^2 \mu _4^6\biggr)\biggr]
   r_e^8\nonumber\\
&-& 147456 a^2 m^3 \pi ^3 \beta  v_2^3 \mu _3^5 \biggl[360 a
   \pi ^4 \beta ^2+m \biggl(10 a m \mu _1^6-5 a \left(m \mu
   _4^3+12 \pi ^2 \beta \right) \mu _1^3\nonumber\\
   &+& 9 m \left(a \mu +a \mu
   _2-\pi \right) \mu _3^5 \nonumber\\
&+& 30 a \pi ^2 \beta  \mu
   _4^3\biggr)\biggr] r_e^7+4096 m^3 v_2^3 \biggl[100 a^2 m^3
   \left(\pi -2 a \mu _5\right) \mu _1^{12}+200 a^2 m^2 \biggl(2
   a \mu _2 \left(m \mu _4^3-12 \pi ^2 \beta \right) \nonumber\\
&+& \pi
   \left(24 a \pi  \beta  \mu _5-m \mu _4^3\right)\biggr) \mu
   _1^9+20 a m \biggl(-54 a^2 m^2 \mu _2^2 \mu _3^5-9 m^2
   \left(\pi  (a \mu +\pi )+a (\pi -2 a \mu ) \mu _5\right) \mu
   _3^5 \nonumber\\
&+& 18 a \pi  \mu _2 \left(3 m^2 \mu _3^5+20 a \pi  \beta
   \left(6 \pi ^2 \beta -m \mu _4^3\right)\right)+5 a \biggl(m^2
   (\pi -2 a \mu ) \mu _4^6+12 m \pi ^2 \beta  (4 a \mu +\pi )
   \mu _4^3 \nonumber\\
&+& 72 \pi ^4 \beta ^2 \left(-3 a \mu -5 a \mu _5+\pi
   \right)\biggr)\biggr) \mu _1^6+60 a \biggl(216 a^2 m^2 \pi ^2
   \beta  \mu _2^2 \mu _3^5 \nonumber\\
&-& 6 a \mu _2 \left(m^2 \left(m (\pi -2
   a \mu ) \mu _4^3+24 \pi ^2 \beta  (a \mu +\pi )\right) \mu
   _3^5+160 a \pi ^4 \beta ^2 \left(3 \pi ^2 \beta -m \mu
   _4^3\right)\right) \nonumber\\
&+& \pi  \biggl(3 m^2 \left(36 a \pi ^2 \beta
   \mu +(\pi -2 a \mu ) \left(m \mu _4^3+12 a \pi  \beta  \mu
   _5\right)\right) \mu _3^5+20 a \pi  \beta  \biggl(-m^2 (\pi -2
   a \mu ) \mu _4^6 \nonumber\\
&-& 6 m \pi ^2 \beta  (6 a \mu +\pi ) \mu
   _4^3+72 a \pi ^4 \beta ^2 \left(\mu +\mu
   _5\right)\biggr)\biggr)\biggr) \mu _1^3 \nonumber\\
   &+&9 \biggl(9 \left(16 m
   \pi ^5 \beta ^2 v_2 a^3+m^3 (\pi -2 a \mu ) \left(\pi -2 a
   \mu _2\right){}^2\right) \mu _3^{10} \nonumber\\
&-& 120 a m \pi ^2 \beta
   \biggl(60 a^2 \pi ^2 \beta  \mu _2^2+2 a \left(-m (\pi -2 a
   \mu ) \mu _4^3-30 \pi ^3 \beta \right) \mu _2+6 \pi ^3 \beta
   (a \mu +2 \pi ) \nonumber\\
&+& \pi  (\pi -2 a \mu ) \left(m \mu _4^3+6 a \pi
    \beta  \mu _5\right)\biggr) \mu _3^5\nonumber\\
    &+&400 a^2 \pi ^4 \beta ^2
   \mu _4^3 \left(m (\pi -2 a \mu ) \mu _4^3+12 \pi ^2 \beta
   \left(2 a \mu -4 a \mu _2+\pi \right)\right)\biggr)\biggr]
   r_e^6 \nonumber\\
&+& 131072 a m^2 \pi ^2 \beta  v_2^4 \biggl[100 a^2 m^3 \mu
   _1^{12}-100 a^2 m^2 \left(m \mu _4^3+12 \pi ^2 \beta \right)
   \mu _1^9 \nonumber\\
&-& 180 a m \biggl(m^2 \left(a \mu -3 a \mu _2+\pi
   \right) \mu _3^5-10 a \pi ^2 \beta  \left(m \mu _4^3+4 \pi ^2
   \beta \right)\biggr) \mu _1^6\nonumber\\
&+& 90 a m \left(m \mu _3^5 \left(m
   (\pi -2 a \mu ) \mu _4^3+12 \pi ^2 \beta  \left(4 a \mu -6 a
   \mu _2+\pi \right)\right)-160 a \pi ^4 \beta ^2 \mu
   _4^3\right) \mu _1^3\nonumber\\
&+& 81 m^3 (\pi -2 a \mu ) \left(\pi -2 a
   \mu _2\right) \mu _3^{10}+43200 a^2 \pi ^6 \beta ^3 \mu
   _4^3\nonumber\\
&-& 540 a m \pi ^2 \beta  \mu _3^5 \left(m (\pi -2 a \mu )
   \mu _4^3+12 \pi ^2 \beta  \left(a \mu -5 a \mu _2+2 \pi
   \right)\right)\biggr] r_e^5\nonumber\\
&-& 9437184 m^2 \pi ^3 \beta ^2 v_2^4
   \biggl[10 a m^2 \left(a \Big(3 a \pi  v_2 \mu _3^5-30 a \mu
   _2^2+30 \pi  \mu _2-5 (\pi -2 a \mu ) \mu _5\right)\nonumber \\
&-&5 \pi  (a
   \mu +\pi )\Big) \mu _1^6
\nonumber\\
&+& 10 a m \biggl(360 a^2 \pi ^2 \beta
   \mu _2^2-10 a \left(m (\pi -2 a \mu ) \mu _4^3+24 \pi ^2
   \beta  (a \mu +\pi )\right) \mu _2 \nonumber\\
&+& \pi  \left(-36 a^2 \pi ^2
   \beta  v_2 \mu _3^5+180 a \pi ^2 \beta  \mu +5 (\pi -2 a \mu
   ) \left(m \mu _4^3+12 a \pi  \beta  \mu
   _5\right)\right)\biggr) \mu _1^3 \nonumber\\
&+& 3 \biggl(-60 a^2 \left(80 a
   \pi ^4 \beta ^2-m^2 (\pi -2 a \mu ) \mu _3^5\right) \mu
   _2^2+20 a \pi  \biggl(60 a \pi ^3 (2 a \mu +3 \pi ) \beta ^2 \nonumber\\
&+& m (\pi -2 a \mu ) \left(10 a \pi  \beta  \mu _4^3-3 m \mu
   _3^5\right)\biggr) \mu _2 \nonumber\\
   &+&\pi  \biggl(3 a^2 m^2 (2 a \mu -\pi
   ) v_2 \mu _3^{10}+15 \pi  \left(40 \pi ^3 \beta ^2 v_2
   a^3+m^2 (\pi -2 a \mu )\right) \mu _3^5 \nonumber\\
&-& 100 a \pi ^2 \beta
   \left(3 \pi ^2 \beta  (6 a \mu +\pi )+(\pi -2 a \mu ) \left(m
   \mu _4^3+6 a \pi  \beta  \mu
   _5\right)\right)\biggr)\biggr)\biggr] r_e^4 \nonumber\\
&+& 1509949440 a m
   \pi ^5 \beta ^3 v_2^5 \biggl[10 a m^2 \left(a \mu -3 a \mu
   _2+\pi \right) \mu _1^6+5 a m \biggl(-m (\pi -2 a \mu ) \mu
   _4^3\nonumber\\
&-& 12 \pi ^2 \beta  \left(4 a \mu -6 a \mu _2+\pi
   \right)\biggr) \mu _1^3- 9 m^2 (\pi -2 a \mu ) \left(\pi -2 a
   \mu _2\right) \mu _3^5 \nonumber\\
&+& 30 a m \pi ^2 \beta  (\pi -2 a \mu )
   \mu _4^3+360 a \pi ^4 \beta ^2 \left(2 a \mu -4 a \mu _2+\pi
   \right)\biggr] r_e^3\nonumber\\
&+& 9059696640 \pi ^6 \beta ^4 v_2^5 \biggl[4
   \pi  v_2 \left(240 a \pi ^4 \beta ^2+m \left(5 a m \mu
   _1^6-60 a \pi ^2 \beta  \mu _1^3-3 m (\pi -2 a \mu ) \mu
   _3^5\right)\right) a^2\nonumber\\
&+& 15 m^2 (\pi -2 a \mu ) \left(\pi -2 a
   \mu _2\right){}^2\biggr] r_e^2 + 4348654387200 a m \pi ^8 \beta
   ^5 (\pi -2 a \mu ) v_2^6 \left(\pi -2 a \mu _2\right)
   r_e\nonumber\\
   &+& 34789235097600 a^2 \pi ^{10} \beta ^6 (\pi -2 a \mu )
   v_2^7\biggr\},\nonumber\\
d_6 & = &  m \biggl[9 a m^2 \pi  r_e^4 \mu _3^5-92160 \pi
   ^3 \beta ^2 (\pi - 2 a \mu ) v_2^2 \nonumber\\
   &+& 16 m r_e^2 v_2 \left(10 a m
   \mu _1^6-120 a \pi ^2 \beta  \mu _1^3+9 m (\pi -2 a \mu ) \mu
   _3^5\right)\biggr]^3.
\label{rencouplings}
\end{eqnarray}

%%%%%%%%%%%%%%%%%%%%%%%%%%%%%%%%%%%%%%%%%%%%%%%%%%%%%%%%%%%%%%%%%%%%%%%%%%%%%%%

\end{document}